\documentclass[aps,preprint,nofootinbib,superscriptaddress]{revtex4-1}

\usepackage{amsmath,amssymb,amsfonts}
\usepackage{amsthm}
\usepackage[utf8]{inputenc}
\usepackage{graphicx}
\usepackage{url}
\usepackage[bookmarks,colorlinks,breaklinks]{hyperref}
\hypersetup{linkcolor=blue,citecolor=blue,filecolor=dullmagenta,urlcolor=blue}
\usepackage{bm,bbm}

\newcommand{\udec}{Departamento de F\'{\i}sica, Universidad de Concepci\'{o}n, Casilla 160-C, Concepci\'{o}n, Chile}
\newcommand{\hw}{Department of Mathematics, Heriot-Watt University,Riccarton, Edinburgh EH14 4AS, U.K.}
\newcommand{\maxw}{Maxwell Institute for Mathematical Sciences,\\
The Tait Institute, Edinburgh, U.K.}
\newcommand{\cecs}{Centro de Estudios Cient\'ificos, Casilla 1469, Valdivia, Chile.}
\begin{document}
\begin{flushright}

\baselineskip=12pt

HWM--13--18\\
EMPG--13--18
%\hfill{ }\\
%\today
\end{flushright}
\title{Topological gravity and Wess-Zumino-Witten term}

\author{Patricio Salgado A.}
\email{pasalgad@udec.cl}
\affiliation{\udec}

\author{Patricio Salgado-Rebolledo}
\email{pasalgad@udec.cl}
\affiliation{\udec}
\affiliation{\cecs}
\email{pasalgado@udec.cl}

\author{Omar Valdivia}
\email{ovaldivi@udec.cl}

\affiliation{\udec}
\affiliation{\hw}
\affiliation{\maxw}

\date{\today}
\begin{abstract}
It is shown that the action for topological gravity in even dimensions found
by A. Chamseddine in ref.\cite{Cha90} is, except a multiplicative constant, a
gauged Wess-Zumino-Witten Term.
\end{abstract}

\maketitle

\section{ \textbf{Introduction}}

Lovelock theory of gravity \cite{Lov71}, \cite{Teitelboim:1987zz}, \cite{Mar91} and its very
interesting Chern--Simons subclasses have been the subject of an intensive
study during the last two decades \cite{Cha90}, \cite{Cha89}, \cite{Chamseddine:1989yz}, \cite{Tro99}, \cite{Zan05}.
Chern--Simons gravity theories have been extended by using transgression forms
instead of Chern--Simons forms as actions \cite{Bor03}, \cite{Mor05},
 \cite{Mor06a}, \cite{Iza05}, \cite{Bor05}, \cite{Mor06a},
\cite{Iza06a}, \cite{Iza06c}.  Chern--Simons and Transgression theories of gravity
are valid only in odd-dimensions and in order to have a well defined
even-dimensional theory it would be necessary some kind of dimensional
reduction or compactification.
In ref.\cite{AlvarezGaume:1984dr}, subsequently ref.\cite{Anabalon:2006fj}, \cite{Anabalon:2007dr},
\cite{Anabalon:2008hi} and most recently ref.\cite{Mora:2011sz}, it was pointed out that
Chern--Simons theories are connected with some even-dimensional structures
known as gauged Wess-Zumino-Witten ($\mathbf{gWZW}$) terms.

The connection between this even-dimensional structure and the Chern--Simons
gravity theories suggest that this mechanism could be regarded as an
alternative to compactification or dimensional reduction.

On the other hand, in ref.\cite{Cha90}, Chamseddine constructed topological
actions for gravity in all dimensions. The odd-dimensional theories are based
on the Chern--Simons forms. Even-dimensional theories use, in addition to
the gauge fields, a scalar field $\phi^a$ in the fundamental representation of the gauge group.

In this work it is shown that the action for topological gravity in even
dimensions found by Chamseddine in ref.\cite{Cha90} is a $\mathbf{gWZW}$.

This article is organized as follows. In section \ref{sect2}, we review some aspects of
the topological gravity theory, the so called Stelle-West formalism and of
the $\mathbf{gWZW}$ term. In section \ref{sect3}, it is shown that the action for the
topological gravity studied in ref.\cite{Cha90} corresponds to a $\mathbf{gWZW}$ term. Section \ref{sect4} concludes the work with some comments and
conclusions. The details of some calculations are summarized in an Appendix \ref{app1} and \ref{app2}.

\section{\textbf{Topological gravity, Stelle-West-formalism and the
}$\mathbf{gWZW}$\textbf{ terms}} \label{sect2}

\subsection{\textbf{Topological gravity}}

Some time ago A.H.~Chamseddine constructed actions for topological gravity in
$1+1$ and in $(2n-1)+1$ dimensions \cite{Cha90}, \cite{Cha89}, \cite{Chamseddine:1989yz}.
These actions were constructed from the product of $n$ field strengths,
$F^{ab}$, and a scalar field $\phi^{a}$ in the fundamental representation of the gauge group.

In $\left(  1+1\right)  $-dimensions the action is given by \cite{Chamseddine:1989yz}%
\begin{equation}
\mathrm{S}^{\left(  1+1\right)  }\left[  A,\phi\right]  =k\int_{M_{2}}%
\epsilon_{abc\text{ }}\phi^{a}F^{bc},\label{1}%
\end{equation}
and in $(2n-1)+1$ dimensions the corresponding action can be written in the
form \cite{Chamseddine:1989yz}, \cite{Cha90}
\begin{equation}
\mathrm{S}^{\left(  2n\right)  }\left[  A,\phi\right]  =k\int_{M_{2n}}%
\epsilon_{a_{1}....a_{2n+1}\text{ }}\phi^{a_{1}}F^{a_{2}a_{3}}...F^{a_{2n}%
a_{2n+1}}\text{,}\label{2}%
\end{equation}
where $F^{ab}=dA^{ab}+A^{ac}A_{c}^{\text{ \ }b}$ and $A$ is a one-form gauge
connection. This action was obtained from a Chern--Simons form using a
dimensional reduction method.

In ref.\cite{Merino:2009ya} a related approach to this problem was discussed. In this
reference it was shown that the action $(\ref{2})$ can be obtained from
the $(2n+1)$-dimensional Chern--Simons gravity genuinely invariant under the
Poincar\'{e} group with suitable boundary conditions. Now we will show that
the action $(\ref{2})$ corresponds to a $\mathbf{gWZW}$ term.

\subsection{The \textbf{Stelle-West-formalism}}

The basic idea of the Stelle-West formalism is founded on the non-linear
realizations studied in refs.\cite{Coleman:1969sm}, \cite{Callan:1969sn}, \cite{Volkov:1973vd}. Following
these references, we consider a Lie group $G$ and its stability subgroup $H$.
The Lie group $G$ has $n$ generators. Let us call $\left\{  X_{i}\right\}
_{i=1}^{n-d}$ the generators of $H$. We shall assume that the remaining
generators $\left\{  Y_{l}\right\}  _{l=1}^{d}$\ are chosen so that they form
a representation of $H.$ In other words, the commutator $\left[  X_{i}%
,Y_{l}\right]  $ should be a linear combination of $Y_{l}$ alone. If the
elements of $G/H$ are denoted by $z$, and if the independent fields needed to
parametrise $z$, are denoted by $\phi^{l}$, i.e., the $\phi^{l}$ parametrize
the coset space $G/H$, then a group element $g\in G$ can be uniquely
represented in the form $g=zh$ where $h$ is an element of $H$ and
$z=e^{-\phi^{l}Y_{l}}$.

When $G$ is the group associated to the $\mathrm{AdS}$ Lie algebra $\left[
\mathbf{P}_{a},\mathbf{P}_{b}\right]  =m^{2}J_{ab};$ $\left[  \mathbf{J}%
_{ab},\mathbf{P}_{c}\right]  =\left(  \eta_{bc}\mathbf{P}_{a}-\eta
_{ac}\mathbf{P}_{b}\right)  ;$ $\left[  \mathbf{J}_{ab},\mathbf{J}%
_{cd}\right]  =\left(  \eta_{ac}\mathbf{J}_{bd}-\eta_{bc}\mathbf{J}_{ad}%
+\eta_{bd}\mathbf{J}_{ac}-\eta_{ad}\mathbf{J}_{bc}\right)  $ whose generators
are $\mathbf{P}_{a},\mathbf{J}_{ab}$ and if the subalgebra $H$ is the Lorentz
algebra $SO(3,1)$ whose generators are $\mathbf{J}_{ab}$, then \cite{Ste80}
(see also \cite{Sal02}, \cite{Sal03b}, \cite{Iza04}, \cite{Grignani:1991nj}
\begin{equation}
\frac{1}{2}W^{ab}\mathbf{J}_{ab}+V^{a}\mathbf{P}_{a}=e^{\phi^{c}\mathbf{P}_{c}}\left[  d+\frac{1}{2}\omega^{ab}\mathbf{J}%
_{ab}+e^{a}\mathbf{P}_{a}\right]  e^{-\phi^{c}\mathbf{P}_{c}}.\label{sw39}%
\end{equation}

Using the commutation relation of the $\mathrm{AdS}$ algebra, we find that the
nonlinear fields $V^{a}$ and $W^{ab}$ are given by
\begin{align}
V^{a} &  =e^{a}+\left(  \cosh x-1\right)  \left(  \delta_{b}^{a}-\frac
{\phi_{b}\phi^{a}}{\phi^{2}}\right)  e^{b}\nonumber\\
&  +\frac{\sinh x}{x}D\phi^{a}-\left(  \frac{\sinh x}{x}-1\right)  \left(
\frac{\phi^{c}d\phi_{c}}{\phi^{2}}\phi^{a}\right)  ,\label{sw39'}\\
W^{ab} &  =\omega^{ab}-m^{2}\frac{\sinh x}{x}\left(  \phi^{a}e^{b}-\phi
^{b}e^{a}\right)  \nonumber\\
&  -m^{2}\left[  \phi^{a}D\phi^{b}-\phi^{b}D\phi^{a}\right]  \left(
\frac{\cosh x-1}{x^{2}}\right)  .\label{sw39''}%
\end{align}
with $x=m\left(  \phi^{a}\phi_{a}\right)  ^{1/2}=m\phi$. 

Taking the limit $m\rightarrow0$ in the commutation relation of the $\mathrm{AdS}$ Lie
algebra and in $(\ref{sw39'})$ and $(\ref{sw39''})$ we find that the $\mathrm{AdS}$ Lie
algebra takes the form of the Poincare Lie algebra where now the nonlinear
fields are given by
\begin{equation}
V^{a}=e^{a}+D\phi^{a};\text{ \ }W^{ab}=\omega^{ab}.
\end{equation}

\subsection{\textbf{The gauged Wess-Zumino-Witten Term}}

\subsubsection{\textbf{Chern Simons Form and WZ Term }}

Consider the gauge transformed field \cite{Bertlmann:1996xk}
\begin{equation}
A^{g}=g^{-1}Ag+g^{-1}dg=g^{-1}\left(  A+\mathcal{V}\right)  g\text{ \ }
\label{wz1}%
\end{equation}
with $\mathcal{V}=dgg^{-1\text{ }}$ and the transformed curvature
\begin{equation}
F^{g}=dA^{g}+\left(  A^{g}\right)  ^{2}=g^{-1}Fg \label{wz2}%
\end{equation}
where $g=g(x)$ denotes the gauge element. Let us choose a homotopy such as
\begin{equation}
A_{t}^{g}=tg^{-1}Ag+g^{-1}dg=g^{-1}\left(  A_{t}+\mathcal{V}\right)  g
\label{wz3}%
\end{equation}
with $A_{t}=tA$, $t\in\left[  0,1\right]$.
The corresponding homotopic curvature is
\begin{equation}
F_{t}^{g}=dA_{t}^{g}+\left(  A_{t}^{g}\right)  ^{2}=g^{-1}F_{t}g\label{wz4}%
\end{equation}
with%
\begin{equation}
F_{t}=dA_{t}+A_{t}^{2}=tF+(t^{2}-t)A^{2}\label{wz5}%
\end{equation}
Both homotopies $(\ref{wz3})$ and $(\ref{wz4})$ interpolate continuously
between $A_{t=1}^{g}=A^{g},$ \ $F_{t=1}^{g}=F^{g}$ and $A_{t=0}^{g}%
=g^{-1}dg=g^{-1}\mathcal{V}g,$ \ $F_{t=0}^{g}=0.$ (See Appendix \ref{app2}.)

Applying the Cartan homotopy formula $(\ref{b9})$ to a Chern-Simons
form$(\ref{b14})$ containing the homotopies $(\ref{wz3})$ and $(\ref{wz4})$,
we have
\begin{equation}
Q_{2n+1}(A_{1}^{g},F_{1}^{g})-Q_{2n+1}(A_{0}^{g},F_{0}^{g})=\left(
k_{01}d+dk_{01}\right)  Q_{2n+1}(A_{t}^{g},F_{t}^{g})\label{wz6}%
\end{equation}%
\begin{equation}
Q_{2n+1}(A^{g},F^{g})-Q_{2n+1}(g^{-1}dg,0)=\left(  k_{01}d+dk_{01}\right)
Q_{2n+1}(A_{t}^{g},F_{t}^{g})\label{wz7}%
\end{equation}
From $(\ref{b14})$ we can see that the gauge transformed Chern-Simons term is
given by \cite{Bertlmann:1996xk}%

\begin{equation}
Q_{2n+1}(A^{g},F^{g})=\left(  n+1\right)  \int_{0}^{1}dt\left\langle
A^{g}\left(  \hat{F}_{t}^{g}\right)  ^{n}\right\rangle \label{wz8}%
\end{equation}
where
\begin{align}
\hat{F}_{t}^{g} &  =tF^{g}+(t^{2}-t)\left(  A^{g}\right)  ^{2}=g^{-1}\hat
{F}_{t}g\nonumber\\
\hat{F}_{t} &  =tF+(t^{2}-t)\left(  A+\mathcal{V}\right)  ^{2}\label{wz9}%
\end{align}
so that
\begin{align}
Q_{2n+1}(A^{g},F^{g}) &  =\left(  n+1\right)  \int_{0}^{1}dt\left\langle
g^{-1}\left(  A+\mathcal{V}\right)  g\left[  g^{-1}\hat{F}_{t}g\right]
^{n}\right\rangle \label{wz10}\\
&  =\left(  n+1\right)  \int_{0}^{1}dt\left\langle g^{-1}\left(
A+\mathcal{V}\right)  \hat{F}_{t}^{n}g\right\rangle =Q_{2n+1}(A+\mathcal{V}%
,F)\nonumber
\end{align}
Analogously, from $(\ref{wz8})$ we can see that%
\begin{equation}
Q_{2n+1}(A_{t}^{g},F_{t}^{g})=\left(  n+1\right)  \int_{0}^{1}ds\left\langle
A_{t}^{g}\left[  \left(  F_{t}^{g}\right)  _{s}\right]  ^{n}\right\rangle
\label{wz11}%
\end{equation}
with
\begin{align}
\left(  F_{t}^{g}\right)  _{s} &  =sF_{t}^{g}+(s^{2}-s)\left(  A_{t}%
^{g}\right)  ^{2}=g^{-1}F_{ts}g\label{wz12}\\
F_{ts} &  =sF_{t}+(s^{2}-s)\left(  A_{t}+\mathcal{V}\right)  ^{2}%
\end{align}
so that
\begin{equation}
Q_{2n+1}(A_{t}^{g},F_{t}^{g})=\left(  n+1\right)  \int_{0}^{1}ds\left\langle
\left(  A_{t}+\mathcal{V}\right)  F_{ts}^{n}\right\rangle =Q_{2n+1}%
(A_{t}+\mathcal{V},F_{t})\label{wz13}%
\end{equation}

To calculate $\left(  k_{01}d+dk_{01}\right)  Q_{2n+1}(A_{t}^{g},F_{t}^{g})$,
remember that the results $(\ref{b13})$ \ $dQ_{2n+1}(A,F)=\left\langle
F^{n+1}\right\rangle $ can be generalized to
\begin{equation}
dQ_{2n+1}(A_{t}^{g},F_{t}^{g})=\left\langle g^{-1}F_{t}^{n+1}g\right\rangle
=\left\langle F_{t}^{n+1}\right\rangle \label{wz14}%
\end{equation}
so that%
\begin{align}
k_{01}dQ_{2n+1}(A_{t}^{g},F_{t}^{g}) &  =k_{01}\left\langle F_{t}%
^{n+1}\right\rangle =\int_{0}^{1}l_{t}\left\langle F_{t}^{n+1}\right\rangle
\nonumber\\
&  =\left(  n+1\right)  \int_{0}^{1}dt\left\langle AF_{t}^{n}\right\rangle
=Q_{2n+1}(A,F)\label{wz15}%
\end{align}

Defining the $2n$-form
\begin{equation}
\alpha_{2n}=k_{01}Q_{2n+1}(A_{t}^{g},F_{t}^{g})=k_{01}Q_{2n+1}(A_{t}%
+\mathcal{V},F_{t})\label{wz16}%
\end{equation}
we find that the Cartan homotopy formula $(\ref{wz7})$ takes the form
\begin{equation}
Q_{2n+1}(A^{g},F^{g})=Q_{2n+1}(A,F)+Q_{2n+1}(\mathcal{V},0)+d\alpha
_{2n}\label{wz18}%
\end{equation}
The second term in the r.h.s. of $(\ref{wz18})$ corresponds to the so called
Wess-Zumino term and since it represents a winding number, it will be
total derivative, unless $G$ has non-trivial homotopy group $\pi_{3}\left(
G\right)  $ and large gauge transformations are performed.

We now consider the terms $Q_{2n+1}(\mathcal{V},0)$ and $\alpha_{2n}$. From $(\ref{wz10})$ with $A=0$, $F=0$, we find
\begin{equation}
Q_{2n+1}(\mathcal{V},0)=\left(  n+1\right)  \int_{0}^{1}dt\left\langle
\mathcal{V}F_{t}^{n}\right\rangle
\end{equation}
where $\hat{F}_{t}=t(t-1)\mathcal{V}^{2}=-t(1-t)\mathcal{V}^{2},$ so that
$\hat{F}_{t}^{n}=\left(  -1\right)  ^{n}t^{n}(1-t)^{n}\mathcal{V}^{2n}$ \ and
therefore \cite{Bertlmann:1996xk}%
\begin{equation}
Q_{2n+1}(\mathcal{V},0)=\left(  -1\right)  ^{n}\frac{n!\left(  n+1\right)
!}{\left(  2n+1\right)  !}\left\langle \mathcal{V}^{2n+1}\right\rangle
\label{21'}%
\end{equation}
which corresponds to the generalization of the Wess-Zumino term.
\subsubsection{\textbf{Cartan homotopy formula and transgression form}}
Applying the Cartan homotopy formula $(\ref{b9})$ to the Chern-Simons form
$(\ref{b14})$ we have \cite{Bertlmann:1996xk}%
\begin{equation}
k_{01}dQ_{2n+1}(A_{t},F_{t})=Q_{2n+1}(A_{1},F_{1})-Q_{2n+1}(A_{0}%
,F_{0})-dk_{01}Q_{2n+1}(A_{t},F_{t})\label{26''}%
\end{equation}
where now $A_{t}=A_{0}+t\left(  A_{1}-A_{0}\right)$. From $(\ref{26''})$ we
can see that%
\begin{equation}
Q_{2n+1}(A_{1},A_{0})=Q_{2n+1}(A_{1},F_{1})-Q_{2n+1}(A_{0},F_{0})-d\left[
k_{01}Q_{2n+1}(A_{t},F_{t})\right]  \label{28'}%
\end{equation}
where
\begin{equation}
Q_{2n+1}(A_{1},A_{0})=k_{01}\left\langle F_{t}^{n+1}\right\rangle
=(n+1)\int_{0}^{1}dt\left\langle \theta F_{t}^{n}\right\rangle \label{27'}%
\end{equation}
is known as the transgression form.

Defining
\begin{equation}
\mathrm{B}_{2n}=k_{01}dQ_{2n+1}(A_{t},F_{t})\label{27''}%
\end{equation}
we find that $(\ref{28'})$ takes the form%
\begin{equation}
Q_{2n+1}(A_{1},A_{0})=Q_{2n+1}(A_{1},F_{1})-Q_{2n+1}(A_{0},F_{0}%
)-dB_{2n}(A_{1},A_{0})\label{32'}%
\end{equation}
where%
\begin{equation}
B_{2n}(A_{1},A_{0})=n(n+1)\int_{0}^{1}dt\int_{0}^{t}ds\left\langle \left(
A_{1}-A_{0}\right)  A_{0}F_{st}^{n-1}\right\rangle \label{31'}%
\end{equation}
with $F_{ts}=sF_{t}+s(s-1)A_{t}^{2},$ $A_{ts}=sA_{t}=sA_{0}+st\left(
A_{1}-A_{0}\right)  ,$ $A_{t}=A_{0}+t\left(  A_{1}-A_{0}\right)$.
\subsubsection{\textbf{Gauged Wess-Zumino-Witten Term}}

If $A_{1}$ is related to $A_{0}$ by a gauge transformation and if $A_{1}%
,A_{0}$ are denoted by $A^{g},A$ respectively, we can write
\begin{equation}
Q_{2n+1}(A^{g},A)=Q_{2n+1}(A^{g},F^{g})-Q_{2n+1}(A,F)-dB_{2n}(A^{g},A)
\label{wzw3}%
\end{equation}
From $(\ref{wz18})$ and $\left(  \ref{wzw3}\right)  $ we can see that the
transgression form for two gauge equivalent connections correspond to the
$\mathbf{gWZW}$ term
\begin{equation}
Q_{2n+1}(A^{g},A)=Q_{2n+1}(\mathcal{V},0)+d\alpha_{2n}-dB_{2n}(A^{g}%
,A).\label{wzw5}%
\end{equation}
In the particular case $n=1$, i.e., in the $(2+1)-$dimensional case, we have
that $\alpha_{2}$ takes the form%
\begin{align}
\alpha_{2} &  =\int_{0}^{1}l_{t}Q_{2n+1}(A_{t}^{g},F_{t}^{g})=\int_{0}%
^{1}l_{t}\left\langle \left(  A_{t}+\mathcal{V}\right)  \left(  F_{t}-\frac
{1}{3}A_{t}^{2}\right)  \right\rangle \nonumber\\
&  =-\left\langle \int_{0}^{1}dt\left(  tA+\mathcal{V}\right)  \frac{\partial
A_{t}}{\partial t}\right\rangle =-\left\langle \mathcal{V}A\right\rangle
,\label{wzw6}%
\end{align}
where we have used $(\ref{b3})$ and $(\ref{b4})$. On the other
hand, from equation $(\ref{31'})$ we see that $B_{2}(A^{g},A)$ is given
by
\begin{equation}
\mathrm{B}_{2}(A^{Z},A)=\left\langle A^{g}A\right\rangle .\label{wzw7}%
\end{equation}
From $(\ref{21'})$, $(\ref{wzw6})$ and $(\ref{wzw7})$ we
can see that the $gWZW$ term in $2+1$ dimensions can be written as \cite{Mora:2011sz}%
\begin{equation}
Q_{2n+1}(A^{g},A)=-\frac{1}{3}\left\langle \mathcal{V}^{3}\right\rangle
-d\left(  \left\langle \mathcal{V}A+A^{g}A\right\rangle \right)  .\label{wzw8}%
\end{equation}

\section{\textbf{Topological gravity as a gauged Wess-Zumino-Witten Term}} \label{sect3}
In this section we show that even-dimensional topological gravity is a $\mathbf{gWZW}$ term.
\subsection{\textbf{Topological Gravity in }$\left(  1+1\right)
$\textbf{-dimensions}}
From $(\ref{sw39})$ we can see that the nonlinear and the
linear connection, $A^{Z}$ and $A=e+\omega$ respectively, are related by a gauge transformation given by
\begin{equation}
A^{Z}=z^{-1}\left(  d+A\right)  z,\label{top1}%
\end{equation}
where $z=e^{-\phi^{a}P_{a}}$ and $A^{Z}=\frac{1}{2}W^{ab}J_{ab}+V^{a}%
P_{a}=V+W$. This means that the linear and nonlinear curvatures $F^{Z}$ and
$F$ are related by
\begin{equation}
F^{Z}=z^{-1}Fz.\label{top2}%
\end{equation}

In the $(2+1)-$dimensional case, the only non-vanishing component is given
by
\[
\left\langle J_{ab}P_{c}\right\rangle =\epsilon_{abc}%
\]
so that the Wess-Zumino term $(\ref{21'})$ vanishes%
\[
Q_{2+1}(\mathcal{V},0)=-\frac{1}{3}\left\langle \mathcal{V}^{3}\right\rangle
=0.
\]
Hence, the $\mathbf{gWZW}$ term $(\ref{wzw8})$ takes the form%
\begin{equation}
Q_{3}(A^{Z},A)=d\left(  \alpha_{2}-B_{2}(A^{Z},A)\right)  =-d\left(
\left\langle \mathcal{V}A+A^{z}A\right\rangle \right)  ,\label{wzw}%
\end{equation}
and defines a Lagrangian in $\left(  1+1\right)  \mathbf{-}$dimensions.
Consider first the term $\left\langle \mathcal{V}A\right\rangle $%
\begin{align*}
\left\langle \mathcal{V}A\right\rangle  &  =\left\langle \mathcal{V}\left(
e+\omega\right)  \right\rangle =-\left\langle \left(  e^{a}P_{a}+\frac{1}%
{2}\omega^{ab}J_{ab}\right)  d\phi^{c}P_{c}\right\rangle \\
&  =-\frac{1}{2}\omega^{ab}d\phi^{c}\left\langle J_{ab}P_{c}\right\rangle \\
&  =-\frac{1}{2}\epsilon_{abc}\omega^{ab}d\phi^{c}%
\end{align*}
which can be rewritten as%
\begin{align}
\left\langle \mathcal{V}A\right\rangle  &  =-\varepsilon_{abc}d\omega^{ab}%
\phi^{c}+d\left(  \frac{1}{2}\epsilon_{abc}\omega^{ab}\phi^{c}\right)
+\frac{1}{2}\epsilon_{abc}\omega^{ab}d\phi^{c}\nonumber\\
&  =-\varepsilon_{abc}\left(  d\omega^{ab}\phi^{c}+\frac{1}{2}\omega
^{ab}\omega^{c}{}_{d}\phi^{d}\right)  +\frac{1}{2}\epsilon_{abc}\omega
^{ab}D\phi^{c}+d\left(  \frac{1}{2}\epsilon_{abc}\omega^{ab}\phi^{c}\right)
\nonumber\\
&  =-\varepsilon_{abc}R^{ab}\phi^{c}+\frac{1}{2}\epsilon_{abc}\omega^{ab}%
D\phi^{c}+d\left(  \frac{1}{2}\epsilon_{abc}\omega^{ab}\phi^{c}\right)
\label{top6'}%
\end{align}
where we have used the identity $\varepsilon_{abc}\omega^{ab}%
\omega^{c}{}_{d}\phi^{d}=2\varepsilon_{abc}\omega^{a}{}_{d}\omega^{db}{}%
\phi^{c}$. On the other hand, the term $\left\langle
A^{z}A\right\rangle $ is given by%
\begin{equation}
\left\langle A^{z}A\right\rangle =\left\langle \left(  A^{z}-A\right)
A\right\rangle =\left\langle D\phi\left(  e+\omega\right)  \right\rangle
=-\frac{1}{2}\varepsilon_{abc}\omega^{ab}D\phi^{c}.\label{aza}%
\end{equation}
Substituting $(\ref{aza})$ and $(\ref{top6'})$ in
$(\ref{wzw})$ we obtain%
\[
Q_{3}(A^{Z},A)=d\left[  \epsilon_{abc}R^{ab}\phi^{c}-d\left(  \frac{1}%
{2}\epsilon_{abc}\omega^{ab}\phi^{c}\right)  \right]  =d\left(  \epsilon
_{abc}R^{ab}\phi^{c}\right)
\]
which proves that the action for Topological gravity in
(1+1)-dimensions, found in ref.\cite{Cha90}, 
\cite{Chamseddine:1989yz}, is a $\mathbf{gWZW}$ term given by,
\begin{equation}
\mathrm{S}_{\mathbf{gWZW}}^{(2+1)}\left[  A^{Z},A\right]  =k\int_{M}%
Q_{3}(A^{Z},A)=k\int_{\partial M}\epsilon_{abc}R^{ab}\phi^{c}\label{20}%
\end{equation}

\subsection{\textbf{Topological Gravity in }$2n\mathbf{-}$\textbf{dimensions}}

From $(\ref{wzw5})$ and $(\ref{wz18})$ we can see that%
\begin{equation}
Q_{2n+1}(A^{Z},A)=Q_{2n+1}(V,0)+d\alpha_{2n}-dB_{2n}(A^{Z},A) \label{top7'}%
\end{equation}
Similarly to the three-dimensional case, the only non-vanishing
component of the invariant tensor is
\begin{equation}
\left\langle J_{a_{1}a_{2}}...J_{a_{2n-1}a_{2n}}P_{a_{2n+1}}\right\rangle
=\frac{2^{n}}{n+1}\varepsilon_{a_{1}a_{2}...a_{2n+1}}\label{inv2}%
\end{equation}
so that the Wess-Zumino term $(\ref{21'})$ vanishes%
\[
Q_{2n+1}(V,0)=\left(  -1\right)  ^{n}\frac{n!\left(  n+1\right)  !}{\left(
2n+1\right)  !}\left\langle \left(  dzz^{-1}\right)  ^{2n+1}\right\rangle =0,
\]
and the $\mathbf{gWZW}$ term defines a Lagrangian for a $2n-$dimensional manifold given
by%
\begin{equation}
Q_{2n+1}(A^{Z},A)=d\left(  \alpha_{2n}-B_{2n}(A^{Z},A)\right)  .
\end{equation}
From $(\ref{wz18})$, $d\alpha_{2n}$ can be derived in a straightforward
way
\begin{equation}
d\alpha_{2n}=Q_{2n+1}(A^{Z},F^{Z})-Q_{2n+1}(A,F)\label{top8'}%
\end{equation}
In fact, the term $Q_{2n+1}(A,F)$ corresponding to the Lagrangian for
$(2n+1)$-dimensional Chern-Simons gravity for the one-form connection $A$, is
given by (see Appendix \ref{app1}.)%
\begin{align}
Q_{2n+1}(A,F) &  =\varepsilon_{a_{1}a_{2}...a_{2n+1}}R^{a_{1}a_{2}%
}...R^{a_{2n-1}a_{2n}}e^{a_{2n+1}}\nonumber\\
&  -n(n+1)d\left\{  \int_{0}^{1}dtt^{n}\left\langle R_{t}^{\text{ }n-1}\omega
e\right\rangle \right\}  \label{21}%
\end{align}
where $R_{t}=d\omega+t\omega^{2}.$

If $A^{z}$ and $A$ are given by $A=e^{a}P_{a}+\frac{1}{2}\omega
^{ab}J_{ab}=e+\omega$ and $A^{Z}=V^{a}P_{a}+\frac{1}%
{2}W^{ab}J_{ab}=V+W,$ where $V^{a}=e^{a}+D_{\omega}\phi
^{a}$ and $W^{ab}=\omega^{ab},$ then 
$Q_{2n+1}(A^{Z},F^{Z})$ is given by %
\begin{align*}
Q_{2n+1}(A^{Z},F^{Z}) &  =\varepsilon_{a_{1}...a_{2n+1}}R^{a_{1}a_{2}%
}...R^{a_{2n-1}a_{2n}}V^{a_{2n+1}}\\
&  -n(n+1)d\left\{  \int_{0}^{1}dtt^{n}\left\langle R_{t}^{\text{ }n-1}\omega
V\right\rangle \right\}
\end{align*}%
\begin{align}
Q_{2n+1}(A^{Z},F^{Z}) &  =\varepsilon_{a_{1}a_{2}...a_{2n+1}}R^{a_{1}a_{2}%
}...R^{a_{2n-1}a_{2n}}e^{a_{2n+1}}\nonumber\\
&  +\varepsilon_{a_{1}a_{2}...a_{2n+1}}R^{a_{1}a_{2}}...R^{a_{2n-1}a_{2n}%
}D\phi^{a_{2n+1}}\nonumber\\
&  -n(n+1)d\left\{  \int_{0}^{1}dtt^{n}\left\langle R_{t}^{\text{ }n-1}\omega
e\right\rangle \right\}  \nonumber\\
&  -n(n+1)d\left\{  \int_{0}^{1}dtt^{n}\left\langle R_{t}^{\text{ }n-1}\omega
D\phi\right\rangle \right\}.  \label{22}%
\end{align}
Introducing $(\ref{21})$ and $(\ref{22})$ in $(\ref{top8'})$
we have%
\begin{align}
d\alpha_{2n} &  =\epsilon_{a_{1}...a_{2n+1}}R^{a_{1}a_{2}}...R^{a_{2n-1}%
a_{2n}}D\phi^{a_{2n+1}}\nonumber\\
&  -n(n+1)d\left\{  \int_{0}^{1}dtt^{n}\left\langle R_{t}^{\text{ }n-1}\omega
D\phi\right\rangle \right\}  \label{top9'}%
\end{align}
On the other hand, from equation$(\ref{31'})$ we can see that $B_{2n}%
(A,A^{Z})$ is given by
\begin{equation}
\mathrm{B}_{2n}(A^{Z},A)=n(n+1)\int_{0}^{1}dt\int_{0}^{t}ds\left\langle
\left(  A^{Z}-A\right)  AF_{st}^{n-1}\right\rangle .\label{23}%
\end{equation}
Since $A=e+\omega$, $A^{Z}=A+D\phi,$ $F_{st}=dA_{st}+A_{st}A_{st}$ and
$A_{st}=tA+s\left(  A^{Z}-A\right)  $ we have%
\begin{equation}
F_{st}=tR_{t}+tT_{t}+sD_{t}D\phi\label{24}%
\end{equation}
where $R_{t}=d\omega+t\omega^{2},$ $T_{t}=de+t\left[  \omega,e\right]  $ and
$D_{t}D\phi=d\left(  D\phi\right)  +t\left[  \omega,D\phi\right]  .$
Introducing (\ref{24}) into (\ref{23}) we find
\begin{equation}
\mathrm{B}_{2n}(A^{Z},A)=n(n+1)\int_{0}^{1}dt\int_{0}^{t}ds\left\langle
D\phi\left(  e+\omega\right)  \left(  tR_{t}+tT_{t}+sD_{t}D\phi\right)
^{n-1}\right\rangle. \label{b}%
\end{equation}

Since the only nonvanishing component of the invariant tensor is (\ref{inv2}),
the only nonzero term in (\ref{b}) is $\left\langle D\phi\omega\left(
tR_{t}\right)  ^{n-1}\right\rangle .$ So that
\begin{equation}
\mathrm{B}_{2n}(A^{Z},A)=-n(n+1)\int_{0}^{1}dtt^{n}\left\langle R_{t}^{\text{
}n-1}\omega D\phi\right\rangle \label{27}%
\end{equation}
Introducing (\ref{top9'}) and (\ref{27}) into (\ref{top7'}) we have
\begin{equation}
Q_{2n+1}(A^{Z},A)=\varepsilon_{a_{1}....a_{2n+1}}R^{a_{1}a_{2}}...R^{a_{2n-1}%
a_{2n}}D\phi^{a_{2n+1}}\label{28}%
\end{equation}
using the Bianchi identity $DR^{ab}=0$ we can write%
\begin{equation}
Q_{2n+1}(A^{Z},A)=d\left[  \varepsilon_{a_{1}....a_{2n+1}}R^{a_{1}a_{2}%
}...R^{a_{2n-1}a_{2n}}\phi^{a_{2n+1}}\right]  \label{29}%
\end{equation}
which proves that the action for Topological gravity in $2n-$%
dimensions, found in Ref. \cite{Cha90}, \cite{Chamseddine:1989yz},
is a $\mathbf{gWZW}$ term given by
\begin{equation}
\mathrm{S}_{\mathbf{gWZW}}^{(2n+1)}\left[  A^{Z},A\right]  =k\int_{M_{2n+1}%
}Q_{2n+1}(A^{Z},A)=k\int_{\partial M_{2n+1}}\epsilon_{a_{1}....a_{2n+1}%
}R^{a_{1}a_{2}}...R^{a_{2n-1}a_{2n}}\phi^{a_{2n+1}}\label{30}%
\end{equation}

\section{\textbf{Comments}} \label{sect4}

We have shown in this work that the action for topological gravity in
$2n$-dimensions, introduced in ref.\cite{Cha90}\cite{Chamseddine:1989yz}, is a gauged
Wess-Zumino-Witten term. This means that the $2n$-dimensional topological
gravity is described by the dynamics of the boundary of a $(2n+1)$
Chern-Simons gravity.

The field $\phi^{a}$, which is necessary to construct this type of topological
gravity in even dimensions \cite{Cha90}, is identified by the coset field
associated with non-linear realizations of the Poincare group $ISO(2n,1)$.
This shows a clear geometric interpretation of this field originally
introduced in an "ad-hoc" manner.

This work was supported in part by FONDECYT Grants N$^{0}$ 1130653 and by
Universidad de Concepci\'{o}n through DIUC Grant N$^{0}$ 212.011.056-1.0. Two
of the authors (PSR, OV) were supported by grants from the Comisi\'{o}n
Nacional de Investigaci\'{o}n Cient\'{\i}fica y Tecnol\'{o}gica CONICYT and
from the Universidad de Concepci\'{o}n, Chile. The authors wish to thank S.
Salgado for enlightening discussions.
\appendix
\section{\textbf{Lagrangian for }$(2n+1)$\textbf{-dimensional Chern-Simons
gravity}} \label{app1}

A Chern--Simons form $Q_{2n+1}(A)$ is a differential form defined for a
connection $A$, whose exterior derivative yields a Chern class. Although Chern classes are gauge invariant, the Chern--Simons forms are not; under gauge
transformations they change by a closed form. A transgression form
$Q_{2n+1}(A,\overset{\_}{A})$ on the other hand, is an invariant differential
form whose exterior derivative is the difference of two Chern classes. It
generalizes the Chern--Simons form and has the additional advantage that it is
gauge invariant.

To obtain the Lagrangian for $(2n+1)$-dimensional Chern--Simons gravity we use
the so called Triangle equation \cite{Nak03}
\[
Q_{2n+1}(A,\bar{A})=Q_{2n+1}(A,\tilde{A})-Q_{2n+1}(\bar{A},\tilde{A}%
)-dQ_{2n}(A,\bar{A},\tilde{A})
\]
with $\tilde{A}=0,$ and the method of separation in subspaces. Let us recall
that the method of separation in subspaces consists of the following steps
\cite{Iza06a}, \cite{Iza05}: The first step is to decompose the algebra into subspaces. In
our case $G=V_{1}\oplus V_{2}$, where $V_{1}$ corresponds to the Lorentz
subalgebra generated by $\left\{  \mathbf{J}_{ab}\right\}  $ and $V_{2}$
corresponds to the subspace spanned by $\left\{  \mathbf{P}_{a}\right\}  $.
The second step is to write the connection as a sum of pieces valued in each
subspace. This means $A=a_{1}+a_{2},$ where $a_{1}=\omega$ and $a_{2}=e.$ The
third step is to use the triangular equation with $\tilde{A}=0,$ $\bar
{A}=\omega$ and $A=\omega+e$.

Since $Q_{2n+1}(A,0)=Q_{2n+1}(A)$ and $Q_{2n+1}(\bar{A},0)=Q_{2n+1}(\bar{A}),$
we can write%

\begin{equation}
Q_{2n+1}(A)=\underset{a}{\underbrace{Q_{2n+1}(A,\bar{A})}}+\underset
{b}{\underbrace{Q_{2n+1}(\bar{A},0)}}+\underset{c}{\underbrace{dQ_{2n}%
(A,\bar{A},0)}} \label{a1}%
\end{equation}

$\left(  a\right)  $ To determine the first term we consider%
\begin{equation}
Q_{2n+1}(A,\bar{A})=(n+1)\int_{0}^{1}dt\left\langle \theta F_{t}%
^{n}\right\rangle
\end{equation}
where $\theta=A-\bar{A}=e$, $A_{t}=\bar{A}+t\theta$ and $F_{t}=dA_{t}%
+A_{t}A_{t}=R+tT$, where $R=d\omega+\omega^{2}$ and $T=de+\left[
\omega,e\right]  $. So that%
\begin{equation}
\mathrm{L}_{\mathrm{T}}^{(2n+1)}(A,\bar{A})=(n+1)\int_{0}^{1}dt\left\langle
e\left(  R+tT\right)  ^{n}\right\rangle \label{a2'}%
\end{equation}
Using Newton's binomial theorem and taking into account that the only
component non-zero, of the invariant tensor is given by
\begin{equation}
\left\langle \mathbf{J}_{a_{1}a_{2}}...\mathbf{J}_{a_{2n-1}a_{2n}}%
\mathbf{P}_{a_{2n+1}}\right\rangle =\frac{2^{n}}{n+1}\epsilon_{a_{1}%
...a_{2n+1}}\label{a2''}%
\end{equation}
it is straightforward to see that
\begin{equation}
Q_{2n+1}(A,\bar{A})=(n+1)\int_{0}^{1}dt\left\langle R^{n}e\right\rangle
=\varepsilon_{a_{1}...a_{2n+1}}R^{a_{1}a_{2}}...R^{a_{2n-1}a_{2n}}e^{a_{2n+1}%
}\label{a5}%
\end{equation}

$(b)$  To determine the second term we consider%

\begin{equation}
Q_{2n+1}(\bar{A},0)=(n+1)\int_{0}^{1}dt\left\langle \bar{A}F_{t}%
^{n}\right\rangle \label{a6}%
\end{equation}
where now $A_{t}=t\bar{A}=t\omega$ and $F_{t}=dA_{t}+A_{t}A_{t}=t\left(
d\omega+t\omega^{2}\right)  =tR_{t},$ with $R_{t}=\frac{1}{2}R_{t}%
^{ab}\mathbf{J}_{ab}.$ So that%
\begin{equation}
Q_{2n+1}(\bar{A},0)=(n+1)\int_{0}^{1}dt\left\langle \omega R_{t}^{\text{ }%
n}\right\rangle =0\label{a7}%
\end{equation}
because the only nonzero invariant tensor is given by $\left(  \ref{a2''}%
\right)  $.

$(c)$ To determine the third term we consider
\begin{equation}
Q_{2n}(A,\bar{A},0)=n(n+1)\int_{0}^{1}dt\int_{0}^{t}ds\left\langle \left(
A-\bar{A}\right)  \bar{A}F_{st}^{n-1}\right\rangle =n(n+1)\int_{0}^{1}%
dt\int_{0}^{t}ds\left\langle e\omega F_{st}^{\text{ }n-1}\right\rangle
\label{a8}%
\end{equation}
where $A_{st}=t\omega+se,$ $F_{st}=dA_{st}+A_{st}A_{st}=tR_{t}+sT_{t}$ with
$R_{t}=d\omega+t\omega^{2}$ and $T_{t}=de++t\left[  \omega,e\right]  .$ So
that%
\begin{equation}
Q_{2n}(A,\bar{A},0)=n(n+1)\int_{0}^{1}dt\int_{0}^{t}ds\left\langle
e\omega\left(  tR_{t}+sT_{t}\right)  ^{n-1}\right\rangle .\label{a9}%
\end{equation}
Using Newton's binomial theorem and taking into account that the only
component non-zero, of the invariant tensor is given by $(\ref{a2''})$, we find
\begin{equation}
Q_{2n+1}(A)=\epsilon_{a_{1}...a_{2n+1}}R^{a_{1}a_{2}}...R^{a_{2n-1}a_{2n}%
}e^{a_{2n+1}}-n(n+1)d\left\{  \int_{0}^{1}dtt^{n}\left\langle R_{t}^{\text{
}n-1}\omega e\right\rangle \right\}  \label{a12}%
\end{equation}
If $n=1$ we have
\[
Q_{3}(\bar{A})=\varepsilon_{a_{1}a_{2}a_{3}}R^{a_{1}a_{2}}e^{a_{3}}-\frac
{1}{2}d\left[  \varepsilon_{a_{1}a_{2}a_{3}}\omega^{a_{1}a_{2}}e^{a_{3}%
}\right]  .
\]

\section{\textbf{Cartan homotopy formula }} \label{app2}

We start with the homotopic connection $A_{t}=A_{0}+t(A_{1}%
-A_{0}),$ \ \ \ \ $t\in\left[  0,1\right]  $ and its curvature $F_{t}%
=dA_{t}+A_{t}^{2}$. The homotopy derivation operator $l_{t}$ is defined by%
\begin{equation}
l_{t}A_{t}=0\label{b3}%
\end{equation}%
\begin{equation}
l_{t}F_{t}=d_{t}A_{t}=dt\frac{\partial}{\partial t}A_{t}=dt\left(  A_{1}%
-A_{0}\right)  \label{b4}%
\end{equation}
acting on polynomials in $A_{t}$ y $F_{t}$. It is direct to see that
\begin{equation}
\left(  l_{t}d+dl_{t}\right)  A_{t}=d_{t}A_{t}=dt\frac{\partial}{\partial
t}A_{t}\label{b5}%
\end{equation}%
\begin{equation}
\left(  l_{t}d+dl_{t}\right)  F_{t}=d_{t}F_{t}=dt\frac{\partial}{\partial
t}F_{t}\label{b6}%
\end{equation}
which implies for any polynomial $S$ in $A_{t}$ and $F_{t}$%
\begin{equation}
\left(  l_{t}d+dl_{t}\right)  S(A_{t},F_{t})=d_{t}S(A_{t},F_{t})=dt\frac
{\partial}{\partial t}S(A_{t},F_{t}).\label{b7}%
\end{equation}
Defining the homotopy operator by%
\begin{equation}
k_{01}=\int_{0}^{1}l_{t}\label{b8}%
\end{equation}
i.e., as the $t$-integrated version of the derivation $l_{t},$ we find that
integrating equation (\ref{b7}) with respect to $t$ we arrive at the Cartan
homotopy formula \cite{Bertlmann:1996xk}
\begin{equation}
S(A_{1},F_{1})-S(A_{0},F_{0})=\left(  k_{01}d+dk_{01}\right)  S(A_{t}%
,F_{t}).\label{b9}%
\end{equation}
In the particular case where the arbitrary polynomial $S(A_{t},F_{t})$ is an
invariant polynomial $P_{n+1}(F_{t})=\left\langle F_{t}^{n+1}\right\rangle $
we have $dP_{n+1}(F_{t})=0$ and the Cartan homotopy formula takes the form
\begin{equation}
P_{n+1}(F_{1})-P_{n+1}(F_{0})=dk_{01}P_{n+1}(F_{t}).\label{b10}%
\end{equation}
Since
\begin{equation}
k_{01}P_{n+1}(F_{t})=\int_{0}^{1}l_{t}P_{n+1}(F_{t})=\left(  n+1\right)
\int_{0}^{1}dtP_{n+1}((A_{1}-A_{0}),F_{t}^{n})=Q_{2n+1}(A_{1},A_{0}%
)\label{b10'}%
\end{equation}
we find that the Cartan homotopy formula supplies the Chern--Weil theorem:
\begin{equation}
P_{n+1}(F_{1})-P_{n+1}(F_{0})=dQ_{2n+1}(A_{1},A_{0}).\label{b11}%
\end{equation}

Finally, choosing the case $A_{0}=0,$ $A_{1}=A$ the homotopy operator $k_{01}$
is usually denoted by $k,$ i.e.,
\begin{equation}
k=\int_{0}^{1}l_{t} \label{b12}%
\end{equation}
Applying Cartan homotopy formula to this special case we find%
\begin{equation}
P_{n+1}(F)=dkP_{n+1}(F_{t})=dQ_{2n+1}(A,F)\label{b13}%
\end{equation}
and equation $(\ref{b10'})$ provides the Chern--Simons form%
\begin{equation}
Q_{2n+1}(A,F)=kP_{n+1}(F_{t})=\left(  n+1\right)  \int_{0}^{1}dtP((A,F_{t}%
^{n})=\left(  n+1\right)  \int_{0}^{1}dt\left\langle AF_{t}^{n}\right\rangle
\label{b14}%
\end{equation}
with $A_{t}=tA$ and $F_{t}=dA_{t}+A_{t}^{2}=tF+(t^{2}-t)A^{2}.$

\bibliographystyle{utphys}
\bibliography{biblio2012}

\end{document}